\documentclass[
  preprint,
  5p,times,
  twocolumn,
  twoside,
  number,
  sort&compress
]{elsarticle}

\usepackage[T1]{fontenc}
\usepackage[utf8]{inputenc}
\usepackage{times}
\usepackage[scaled=0.85]{helvet}
\usepackage{graphicx}
\usepackage{multirow}
\usepackage{booktabs,tabularx}
\usepackage[pdftex,colorlinks=false]{hyperref}
\usepackage{microtype}
\usepackage{xspace}

\usepackage{enumitem} \setlist[description]{font=\normalfont\itshape}

\usepackage{ifthen}
\usepackage{xspace}

\newboolean{showcomments}
\setboolean{showcomments}{true}
\ifthenelse{\boolean{showcomments}}
  {\newcommand{\bnote}[2]{
	\fbox{\bfseries\sffamily\scriptsize#1}
    {\sf\small\textit{#2}}
    }
  }
  {\newcommand{\bnote}[2]{}}

\graphicspath{{figures/}}

\newcommand{\tabref}[1]{Table~\ref{tab:#1}}

\newcommand{\commented}[1]{}

\newcommand{\eg}{\emph{e.g.,}\xspace}
\newcommand{\ie}{\emph{i.e.,}\xspace}

\newcommand{\ct}[1]{{\textsf{#1}}\xspace}

\usepackage{alltt}

\usepackage{url}
\makeatletter
\def\url@leostyle{\@ifundefined{selectfont}{\def\UrlFont{\sf}}{\def\UrlFont{\small\sffamily}}}
\makeatother
\journal{Science of Computer Programming}

\begin{document}

\begin{frontmatter}
  \title{A Critical Analysis of String APIs:\\ the Case of Pharo}

  \author[rmod]{Damien Pollet}
  \ead{damien.pollet@inria.fr}

  \author[rmod]{Stéphane Ducasse}
  \ead{stephane.ducasse@inria.fr}

  \address[rmod]{RMoD --- Inria \& Université Lille, France}

  \begin{abstract}
    Most programming languages, besides C, provide a native abstraction for character strings, but string APIs vary widely in size, expressiveness, and subjective convenience across languages.
    In Pharo, while at first glance the API of the String class seems rich, it often feels cumbersome in practice; to improve its usability, we faced the challenge of assessing its design.
    However, we found hardly any guideline about design forces and how they structure the design space, and no comprehensive analysis of the expected string operations and their different variations.
    In this article, we first analyse the Pharo~4 String library, then contrast it with its Haskell, Java, Python, Ruby, and Rust counterparts.
    We harvest criteria to describe a string API, and reflect on features and design tensions.
    This analysis should help language designers in understanding the design space of strings, and will serve as a basis for a future redesign of the string library in Pharo.
  \end{abstract}

\end{frontmatter}

\section{Introduction}
\label{sec:intro}

While strings are among the basic types available in most programming languages, we are not aware of design guidelines, nor of a systematic, structured analysis of the string API design space in the literature.
Instead, features tend to accrete through ad-hoc extension mechanisms, without the desirable coherence.
However, the set of characteristics that good APIs exhibit is generally accepted \cite{Blan08a}; a good API:
\begin{itemize}
\item is easy to learn and memorize,
\item leads to reusable code,
\item is hard to misuse,
\item is easy to extend,
\item is complete.
\end{itemize}
To evolve an understandable API, the maintainer should assess it against these goals.
Note that while orthogonality, regularity and consistency are omitted, they arise from the ease to learn and extend the existing set of operations.
In the case of strings, however, these characteristics are particularly hard to reach, due to the following design constraints.

For a single data type, strings tend to have a large API: in Ruby, the \ct{String} class provides more than 100~methods, in Java more than 60, and Python's \ct{str} around 40.
In Pharo\footnote{Numbers from Pharo~4, but the situation in Pharo~3 is very similar.}, the \ct{String} class alone understands 319~distinct messages, not counting inherited methods.
While a large API is not always a problem \emph{per se}, it shows that strings have many use cases, from concatenation and printing to search-and-replace, parsing, natural or domain-specific languages.
Unfortunately, strings are often abused to eschew proper modeling of structured data, resulting in inadequate serialized representations which encourage a procedural code style\footnote{Much like with Anemic Domain Models, except the string API is complex: \url{http://www.martinfowler.com/bliki/AnemicDomainModel.html}}.
This problem is further compounded by overlapping design tensions:
\begin{description}
\item[Mutability:]
  Strings as values, or as mutable sequences.
\item[Abstraction:]
  Access high-level contents (words, lines, patterns), as opposed to representation (indices in a sequence of characters, or even bytes and encodings).
\item[Orthogonality:]
  Combining variations of abstract operations; for instance, substituting one/several/all occurrences corresponding to an index/character/sequence/pattern, in a case-sensitive/insensitive way.
\end{description}
In previous work, empirical studies focused on detecting non-obvious usability issues with APIs \cite{Styl06a,Styl07a,Picc13a}; for practical advice on how to design better APIs, other works cite guideline inventories built from experience \cite{Beck97a,Cwal05a}.
Joshua Bloch's talk \cite{Bloch06a} lists a number of interesting rules of thumb, but it does not really bridge the gap between abstract methodological advice (e.g. \emph{API design is an art, not a science}) and well-known best practices (e.g. \emph{Avoid long parameter lists}).
Besides the examples set by particular implementations in existing languages like Ruby, Python, or Icon~\cite{Gris96a}, and to the best of our knowledge, we are not aware of string-specific analyses of existing APIs or libraries and their structuring principles.

In this paper, we are not in a position to make definitive, normative design recommendations for a string library; instead, we adopt a descriptive approach and survey the design space to spark discussion around its complexity and towards more understandable, reusable, and robust APIs.
To this end, we study the string libraries of a selection of programming languages, most object-oriented for a comparison basis with Pharo, with Haskell and Rust thrown in for some contrast due to their strong design intents.
We consider these languages to be general purpose and high-level enough that readability, expressivity, and usability are common goals.
However, a caveat: each language comes with its own culture, priorities, and compromises;
we thus have to keep a critical eye and put our findings in the perspective both of the design intent of the studied language, and of our own goals in Pharo.
Similarly, we focus the study on the API of the String class or its equivalent only, and we limit the discussion of related abstractions to their interactions in the string API.
Extending the study to the APIs of other text processing abstractions like streams, regular expressions, or parser combinators at the same level of detail as strings would only make the paper longer.

Section~\ref{sec:problem} shows the problems we face using the current Pharo~4 string library.
In Sections~\ref{sec:patterns} and~\ref{sec:smells}, we identify idioms and smells among the methods provided by Pharo's \ct{String} class.
Section~\ref{sec:ansi} examines the relevant parts of the ANSI Smalltalk standard.
We survey the features expected of a String API in Section~\ref{sec:features}, then existing implementations in several general-purpose languages such as Java, Haskell, Python, Ruby, and Rust in Section~\ref{sec:others}.
Finally, we highlight a few design concerns and takeaways in Section~\ref{sec:takeaways}, before concluding the paper.

\section{Pharo: Symptoms of Organic API Growth}
\label{sec:problem}

As an open-source programming environment whose development branched off from Squeak, Pharo inherits many design decisions from the original Smalltalk-80 library.
However, since the 1980's, that library has grown, and its technical constraints have evolved.
In particular, since Squeak historically focused more on creative and didactic experimentation than software engineering and industrial use, the library has evolved organically more than it was deliberately curated towards a simple and coherent design.

Even though we restrict the scope of the analysis to the \ct{String} class, we face several challenges to identify recurring structures and idioms among its methods, and to understand and classify the underlying design decisions.

\paragraph{Large number of responsibilities}
As explained in Section~\ref{sec:intro}, strings propose a wide, complex range of features.
For example, Pharo's \ct{String} defines a dozen class variables for character and encoding properties.

\paragraph{Large number of methods}
The current Pharo \ct{String} class alone has 319~methods, excluding inherited methods.
However, Pharo supports open-classes: a package can define \emph{extension methods} on classes that belong to another package~\cite{Berg05a,Clif00a};
we therefore exclude extension methods, since they are not part of the core behavior of strings.
Still, this leaves 180~methods defined in the package of \ct{String}.
That large number of methods makes it difficult to explore the code, check for redundancies, or ensure completeness of idioms.

Using the code browser, the developer can group the methods of a class into protocols.
However, since a method can only belong to one protocol, the resulting classification is not always helpful to the user.
For example, it is difficult to know at first sight if a method is related to character case, because there is no dedicated protocol; instead, the case conversion methods are all part of a larger \emph{converting} protocol which bundles conversions to non-string types, representation or encoding conversions, extracting or adding prefixes.

\paragraph{Multiple intertwined behaviors}
Strings provide a complex set of operations for which it is difficult to identify a simple taxonomy.
Consider the interaction between features:
a single operation can be applied to one or multiple elements or the whole string, and can use or return an index, an element, a subset or a subsequence of elements:
\begin{description}
\item[Operations:] insertion, removal, substitution, concatenation or splitting
\item[Scope:] element, pattern occurrence, anchored subsequence
\item[Positions:] explicit indices, intervals, matching queries
\item[Occurrences:] first, last, all, starting from a given one
\end{description}

In Pharo we can replace all occurrences of one character by another one using the \ct{replaceAll:with:} inherited from \ct{SequenceableCollection}, or all occurrences of one character by a subsequence (\ct{copyReplaceAll:with:}).
Like these two messages, some operations will copy the receiver, and some other will change it in place.
This highlights that strings are really mutable collections of characters, rather than pieces of text, and that changing the size of the string requires to copy it.
Finally, replacing only one occurrence is yet another cumbersome message (using \ct{replaceFrom:to:with:startingAt:}).

\begin{tabbing}
\ct{'aaca' replaceAll: \$a with: \$b}
\`{}$\rightarrow$ \ct{'bbcb'} \\

\ct{'aaca' copyReplaceAll: 'a' with: 'bz'}
\`{}$\rightarrow$ \ct{'bzbzcbz'} \\

\ct{'aaca' replaceFrom: 2 to: 3 with: 'bxyz' startingAt: 2}
\`{}$\rightarrow$ \ct{'axya'}
\end{tabbing}

\paragraph{Lack of coherence and completeness}
Besides its inherent complexity, intertwining of behaviors means that, despite the large number of methods, there is still no guarantee that all useful combinations are provided.
Some features are surprisingly absent or unexploited from the basic \ct{String} class.
For instance, string splitting and regular expressions, which are core features in Ruby or Python, have long been third-party extensions in Pharo.
They were only recently integrated, so some methods like \ct{lines}, \ct{substrings:}, or \ct{findTokens:} still rely on ad-hoc implementations.
This reveals refactoring opportunities towards better composition of independent parts.

Moreover, some methods with related behavior and similar names constrain their arguments differently.
For instance, \ct{findTokens:} expects a collection of delimiter characters, but also accepts a single character; however, \ct{findTokens:keep:} lacks that special case.
Perhaps more confusingly, some methods with similar behavior use dissimilar wording: compare the predicates \ct{isAllDigits} and \ct{onlyLetters}, or the conversion methods \ct{asUppercase} and \ct{asLowercase} but \ct{withFirstCharacterDownshifted}.

\paragraph{Impact of immutability}
In some languages such as Java and Python, strings are immutable objects, and their API is designed accordingly.
In Smalltalk, strings historically belong in the collections hierarchy, and therefore are mutable.

In practice, many methods produce a modified copy of their receiver to avoid modifying it in place, but either there is no immediate way to know, or the distinction is made by explicit naming.
For instance, \ct{replaceAll:with:} works in-place, while \ct{copyReplaceAll:with:} does not change its receiver.
Moreover, the VisualWorks implementation supports object immutability, which poses the question of how well the historic API works in the presence of immutable strings.

\paragraph{Duplicated or irrelevant code}
A few methods exhibit code duplication that should be factored out.
For instance, \ct{withBlanksCondensed} and \ct{withSeparatorsCompacted} both deal with repeated whitespace, and \ct{findTokens:} and \ct{findTokens:keep:} closely duplicate their search algorithm.

Similarly, some methods have no senders in the base image, or provide ad-hoc behavior of dubious utility.
For instance, the method comment of \ct{findWordStart:startingAt:} mentions ``HyperCard style searching'' and implements a particular pattern match that is subsumed by a simple regular expression.

\section{Recurring Patterns}
\label{sec:patterns}

We list here the most prominent patterns or idioms we found among the analyzed methods.
Although these patterns are not followed systematically, many of them are actually known idioms that apply to general Smalltalk code, and are clearly related to the ones described by Kent Beck \cite{Beck97a}.
This list is meant more as a support for discussion than a series of precepts to follow.

\begin{figure}[b]
  \includegraphics[width=\linewidth]{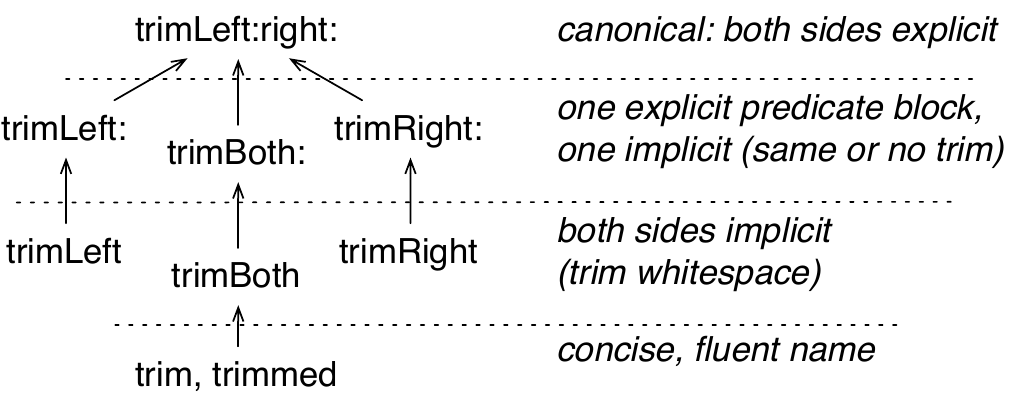}
  \caption{Chains of convenience methods delegating to a single canonical behavior:
    trimming at one or both ends.
    \label{fig:trimming}}
\end{figure}

\paragraph{Layers of convenience}
One of the clearest instances in this study is the group of methods for trimming (Figure~\ref{fig:trimming}).
Trimming a string is removing unwanted characters (usually whitespace) from one or both of its extremities.

The library provides a single canonical implementation that requires two predicates to identify characters to trim at each end of the string.
A first layer of convenience methods eliminates the need for two explicit predicates, either by passing the same one for both ends, or by passing one that disables trimming at one end (\ct{trimBoth:}, \ct{trimLeft:}, and \ct{trimRight:}).
A second layer of convenience methods passes the default predicate that trims whitespace (\ct{trimLeft}, \ct{trimBoth}, and \ct{trimRight}).
Finally, two additional methods provide concise verbs for the most common case: whitespace, both ends (\ct{trim} and \ct{trimmed}, which are synonymous despite the naming).

Convenience methods can also change the result type; the following list shows a few examples of convenience predicates wrapping indexing methods.

\begin{description}
\item[Trimming ends]
  \ct{trim}, \ct{trimmed},
  \ct{trimLeft:right:},\\
  \ct{trimBoth}, \ct{trimBoth:},
  \ct{trimLeft}, \ct{trimleft:},
  \ct{trimRight}, \ct{trimRight:}
\item[Index of character]
  \ct{indexOf:},
  \ct{indexOf:startingAt:},\\
  \ct{indexOf:startingAt:ifAbsent:}
\item[Index of substring]
  \ct{findString:},
  \ct{findString:startingAt:},\\
  \ct{findString:startingAt:caseSensitive:},
  and related predicates \\
  \ct{includesSubstring:},
  \ct{includesSubstring:caseSensitive:}
\item[Macro expansion]
  \ct{expandMacros},
  \ct{expandMacrosWith:} etc.,
  \ct{expandMacrosWithArguments:}
\item[Sort order]
  \ct{compare:},
  \ct{compare:caseSensitive:},\\
  \ct{compare:with:collated:},
  and predicates
  \ct{sameAs:},
  \ct{caseInsensitiveLessOrEqual:}, and
  \ct{caseSensitiveLessOrEqual:}
\item[Spelling correction]
  \ct{correctAgainst:},
  \ct{correctAgainst:continued\-From:},
  \ct{correctAgainstDictionary:continuedFrom:},
  \ct{correct\-AgainstEnumerator:continuedFrom:}
\item[Lines]
  \ct{lines},
  \ct{lineCount},
  \ct{lineNumber:},
  \ct{lineCorrespondingToIndex:},
  \ct{linesDo:},
  \ct{lineIndicesDo:}
\item[Missed opportunity] \ct{substrings} does not delegate to \ct{substrings:}
\end{description}

\noindent
This idiom allows concise code when there is a convention or an appropriate default, without giving up control in other cases.
However, its induced complexity depends on the argument combinations necessary; it then becomes difficult to check all related methods for consistency and completeness.

We propose to broaden and clarify the use of this idiom wherever possible, as it is an indicator of how flexible the canonical methods are, and promotes well-factored convenience methods.
There are several missed opportunities for applying this idiom in \ct{String}: for instance \ct{copyFrom:to:} could have \ct{copyFrom:} (up to the end) and \ct{copyTo:} (from the start) convenience methods.

\paragraph{Pluggable sentinel case}
When iterating over a collection, it is common for the canonical method to expect a block to evaluate for degenerate cases.
This leads to methods that are more akin to control flow, and that let the caller define domain computation in a more general and flexible way.

Methods that follow this idiom typically include either \ct{ifNone:} or \ct{ifAbsent:} in their selector.
For context, in a typical Pharo image as a whole, there are 47~instances of the \ct{ifNone:} pattern, and 266~instances of \ct{ifAbsent:}.
\begin{description}
\item[Index lookup]
  \ct{indexOf:startingAt:ifAbsent:}, \\
  \ct{indexOfSubCollection:startingAt:ifAbsent:}
\end{description}

\noindent
We promote this idiom in all cases where there isn't a clear-cut choice of how to react to degenerate cases.
Indeed, forcing either a sentinel value, a Null Object~\cite{Wool98a}, or an exception on user code forces it to check the result value or catch the exception, then branch to handle special cases.
Instead, by hiding the check, the pluggable sentinel case enables a more confident, direct coding style.
Of course, it is always possible to fall back to either a sentinel, null, or exception, via convenience methods.

\paragraph{Sentinel index value}
When they fail, many index lookup methods return an out-of-bounds index; methods like \ct{copyFrom:to:} handle these sentinel values gracefully.
However, indices resulting from a lookup have two possible conflicting interpretations: either \emph{place of the last match} or \emph{last place examined}.
In the former case, a failed lookup should return zero (since Smalltalk indices are one-based); in the latter case, one past the last valid index signifies that the whole string has been examined.
Unfortunately, both versions coexist:
\begin{tabbing}
\ct{'abc' findString: 'x' startingAt: 1}
\`{}$\rightarrow$ \ct{0} \\
\ct{'abc' findAnySubStr: \#('x' 'y') startingAt: 1}
\`{}$\rightarrow$ \ct{4}
\end{tabbing}
We thus prefer the pluggable sentinel, leaving the choice to user code, possibly via convenience methods.

\begin{description}
\item[Zero index]
  \ct{findSubstring:in:startingAt:matchTable:},
  \ct{findLastOccurrenceOfString:startingAt:},
  \ct{findWordStart:startingAt:},
  \ct{indexOf:startingAt:},
  \ct{indexOfFirstUppercaseCharacter},
  \ct{indexOfWideCharacterFrom:to:},
  \ct{lastSpacePosition},
  \ct{indexOfSubCollection:}
\item[Past the end]
  \ct{findAnySubStr:startingAt:},
  \ct{findCloseParenthesisFor:},
  \ct{findDelimiters:startingAt:}
\end{description}

\paragraph{Iteration or collection}
Some methods generate a number of separate results, accumulating and returning them as a collection.
This results in allocating and building an intermediate collection, which is often unnecessary since the calling code needs to iterate them immediately.
A more general approach is to factor out the iteration as a separate method, and to accumulate the results as a special case only.
A nice example is the group of line-related methods that rely on \ct{lineIndicesDo:}; some even flatten the result to a single value rather than a collection.

\begin{description}
\item[Collection]
  \ct{lines},
  \ct{allRangesOfSubstring:},
  \ct{findTokens:}, \ct{findTokens:keep:}, \ct{findTokens:escapedBy:},
  \ct{substrings}, \ct{substrings:}
\item[Iteration]
  \ct{linesDo:},
  \ct{lineIndicesDo:}
\end{description}

\noindent
In our opinion, this idiom reveals a wider problem with Smalltalk's iteration methods in general, which do not decouple the iteration per se from the choice of result to build --- in fact, collections define a few optimized methods like \ct{select:thenCollect:} to avoid allocating an intermediate collection.
There are many different approaches dealing with abstraction and composeability in the domain of iteration: push or pull values, internal or external iteration, generators, and more recently transducers \cite{Mure96a,transducers}.

\paragraph{Conversion or manipulation}
\ct{String} provides 24~methods whose selector follows the \ct{as\emph{Something}} naming idiom, indicating a change of representation of the value.
Conversely, past participle selectors, e.g. \ct{negated} for numbers, denote a transformation of the value itself, therefore simply returning another value of the same type.
However, this is not strictly followed, leading to naming inconsistencies such as \ct{asUppercase} vs. \ct{capitalized}.

\begin{description}
\item[Type conversions]
  \ct{asByteArray},
  \ct{asByteString},
  \ct{asDate},
  \ct{asDateAndTime},
  \ct{asDuration},
  \ct{asInteger},
  \ct{asOctetString},
  \ct{asSignedInteger},
  \ct{asString},
  \ct{asStringOrText},
  \ct{asSymbol},
  \ct{asTime},
  \ct{asUnsignedInteger},
  \ct{asWideString}
\item[Value transformation or escapement]
  \ct{asCamelCase},
  \ct{asComment},
  \ct{asFourCode},
  \ct{asHTMLString},
  \ct{asHex},
  \ct{asLegalSelector},
  \ct{asLowercase},
  \ct{asPluralBasedOn:},
  \ct{asUncommentedCode},
  \ct{asUppercase} 
\end{description}

\noindent
Past participles read more fluidly, but they do not always make sense, e.g. \ct{commented} suggests adding a comment to the receiver, instead of converting it to one.
Conversely, adopting \ct{as\emph{Something}} naming in all cases would be at the price of some contorted English (\ct{asCapitalized} instead of \ct{capitalized}).

\section{Inconsistencies and Smells}
\label{sec:smells}

Here we report on the strange things we found and that could be fixed or
improved in the short term.

\paragraph{Redundant specializations}
Some methods express a very similar intent, but with slightly differing parameters, constraints, or results.
When possible, user code should be rewritten in terms of a more general approach; for example, many of the pattern-finding methods could be expressed as regular expression matching.

\begin{description}
\item[Substring lookup]
  \ct{findAnySubStr:startingAt:} and \ct{findDelimiters:\-startingAt:} are synonymous if their first argument is a collection of single-character delimiters; the difference is that the former also accepts string delimiters.
\item[Character lookup]
  \ct{indexOfFirstUppercaseCharacter} is redundant with \ct{SequenceableCollection>>findFirst:} with very little performance benefit.
\end{description}

\paragraph{Ad-hoc behavior}
Ad-hoc methods simply provide convenience behavior that is both specific and little used.
Often, the \emph{redundant specialization} also applies.

\begin{description}
\item[Numeric suffix] \ct{numericSuffix} has only one sender in the base Pharo image; conversely, it is the only user of \ct{stemAndNumericSuffix} and \ct{endsWithDigit}; similarly, \ct{endsWithAColon} has only one sender.
\item[Finding text] \ct{findLastOccurrenceOfString:startingAt:} has only one sender, related to code loading; \ct{findWordStart:startingAt:} has no senders.
\item[Find tokens] \ct{findTokens:escapedBy:} has no senders besides tests; \ct{findTokens:includes:} has only one sender, related to email address detection; \ct{findTokens:keep:} only has two senders.
\item[Replace tokens] \ct{copyReplaceTokens:with:} has no senders and is convenience for \ct{copyReplaceAll:with:asTokens:}; redundant with regular expression replacement.
\item[Miscellaneous] \ct{lineCorrespondingToIndex}
\end{description}

\paragraph{Mispackaged or misclassified methods}
There are a couple methods that do not really belong to \ct{String}:
\begin{itemize}
\item
  \ct{asHex} concatenates the literal notation for each character (\eg~\ct{16r6F}) without any separation, producing an ambiguous result; it could be redefined using \ct{flatCollect:}.
\item 
  \ct{indexOfSubCollection:} should be defined in \ct{SequenceableCollection}; also, it is eventually implemented in terms of \ct{findString:}, which handles case, so it is not a simple subsequence lookup.
\end{itemize}
Many ad-hoc or dubious-looking methods with few senders seem to come from the completion engine; the multiple versions and forks of this package have a history of maintenance problems, and it seems that methods that should have been extensions have been included in the core packages.

\paragraph{Misleading names}
Some conversion-like methods are actually encoding or escaping methods: they return another string whose contents match the receiver's, albeit in a different representation (uppercase, lowercase, escaped for comments, as HTML…).

\paragraph{Duplicated code}
Substring testing methods \ct{beginsWithEmpty:caseSensitive:} and \ct{occursInWithEmpty:caseSensitive:} are clearly duplicated: they only differ by a comparison operator.
They are also redundant with the generic \ct{beginsWith:}, except for case-sensitivity.
Moreover, the \ct{--WithEmpty:} part of their selector is confusing; it suggests that argument is supposed to be empty, which makes no sense.
Finally, their uses hint that were probably defined for the completion engine and should be packaged there.

\section{The ANSI Smalltalk Standard}
\label{sec:ansi}

The ANSI standard defines some elements of the Smalltalk language~\cite{ANSI98a}.
It gives the definition \emph{``String literals define objects that represent sequences of characters.''}
However, there are few guidelines helpful with designing a string API.

\begin{figure}[ht]
  \includegraphics[width=\linewidth]{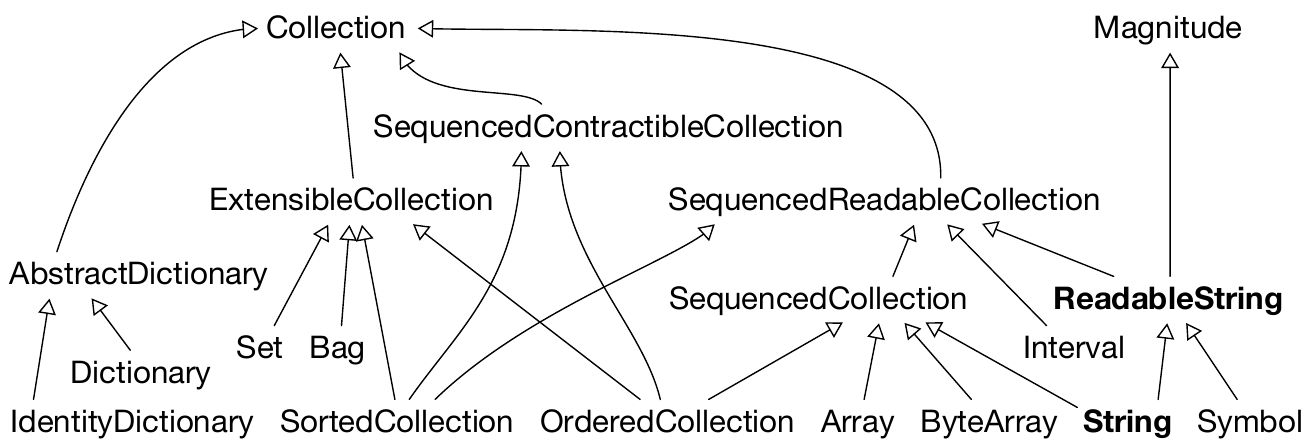}
  \caption{Inheritance of the ANSI Smalltalk protocols.\label{ansi}}
\end{figure}

The ANSI standard defines the \ct{readableString} protocol as conforming to the magnitude protocol (which supports the comparison of entities) and to the \ct{sequencedReadableCollection} protocol, as shown in Figure~\ref{ansi} \cite[section~5.7.10]{ANSI98a}.
We present briefly the protocol \ct{sequencedReadableCollection}.

\paragraph{SequencedReadableCollection}
The \ct{sequencedReadableCollection} protocol conforms to the collection protocol; it provides behavior for reading an ordered collection of objects whose elements can be accessed using external integer keys between one and the number of elements in the collection.
It specifies that the compiler should support the following messages --- we add some of the argument names for clarity:
\begin{description}
\item[Concatenation:]
  \ct{\textbf{,} tail} (the \emph{comma} binary message)
\item[Equality:] \ct{= other}
\item[Element access:]
  \ct{at: index}, \ct{at: index ifAbsent: block},
  \ct{first}, \ct{last},
  \ct{before: element}, \ct{after:},
  \ct{findFirst: block}, \ct{findLast:}
\item[Subsequence access:] \ct{from: startIndex to: stopIndex do: block}
\item[Transforming:] \ct{reverse}
\item[Substitution:]
  \ct{copyReplaceAll: elements with: replacingElements},
  \ct{copyReplaceFrom: startIndex to: stopIndex with: replacingElements},
  \ct{copyReplacing: targetElement withObject: replacingElement},
  \ct{copyReplaceFrom: startIndex to: stopIndex withObject: replacingElement}
\item[Index of element(s):]
  \ct{indexOf: element}, \ct{indexOf:ifAbsent:}, \\
  \ct{indexOfSubCollection:startingAt:}, \\
  \ct{indexOfSubCollection:startingAt:ifAbsent:}
\item[Copy:] \ct{copyFrom: startIndex to: lastIndex}, \ct{copyWith: element}, \ct{copyWithout:}
\item[Iteration:] \ct{do:}, \ct{from:to:keysAndValuesDo:}, \ct{keysAndValuesDo:}, \ct{reverseDo:}, \ct{with:do:}
\end{description}
Many operations require explicit indices that have to be obtained first, making the API not very fluid in practice.
Moreover, the naming is often obscure: for example, \ct{copyWith:} copies the receiver, and \emph{appends} its argument to it.

\paragraph{ReadableString}
This protocol provides messages for string operations such as copying, comparing, replacing, converting, indexing, and matching.
All objects that conform to the \ct{readableString} protocol are comparable.
The copying messages inherited from the \ct{sequencedReadableCollection} protocol keep the same behavior. 
Here is the list of messages:
\begin{description}
\item[Concatenation:] \ct{\textbf{,}} (comma)
\item[Comparing:] \ct{<}, \ct{<=}, \ct{>}, \ct{>=}
\item[Converting:] \ct{asLowercase}, \ct{asString}, \ct{asSymbol}, \ct{asUppercase}
\item[Substituing:]
  \ct{copyReplaceAll:with:},
  \ct{copyReplaceFrom:to:with:},
  \ct{copyReplacing:withObject:},
  \ct{copyWith:}
\item[Subsequence access:] \ct{subStrings: separatorCharacters}
\item[Testing:] \ct{sameAs:}
\end{description}

\paragraph{Analysis and ANSI Compliance}
Indices are omnipresent, and very few names are specific to strings as opposed to collections, which makes the protocol feel shallow, low-level and implementation revealing.
In particular, because the underlying design is stateful, the \ct{copyReplace*} messages have to explicitly reveal that they do not modify their receiver through cumbersome names.
In a better design, naming would encourage using safe operations over unsafe ones.

We believe that the value added by complying with the ANSI standard is shallow.
Indeed, the standard has not been updated to account for evolutions such as immutability, and it does not help building a fluent, modern library.
ANSI should not be followed for the design of a modern String library.

\section{An Overview of Expected String Features}
\label{sec:features}

Different languages do not provide the exact same feature set\footnote{They can even rely on specific syntax, like Ruby's string interpolation.}, or the same level of convenience or generality.
However, comparing various programming languages, we can identify the main behavioral aspects of strings.
Note that these aspects overlap:
for instance, transposing a string to upper-case involves substitution, and can be performed in place or return a new string;
splitting requires locating separators and extracting parts as smaller strings, and is a form of parsing.

\paragraph{Extracting}
Locating or extracting parts of a string can be supported by specifying either explicit indices, or by matching contents with various levels of expressiveness: ad-hoc pattern, character ranges, regular expressions.

\paragraph{Splitting}
Splitting strings into chunks is the basis of simple parsing and string manipulation techniques, like counting words or lines in text.
To be useful, splitting often needs to account for representation idiosyncrasies like which characters count as word separators or the different carriage return conventions.

\paragraph{Merging}
The reverse of splitting is merging several strings into one, either by concatenation of two strings, or by joining a collection of strings one after another, possibly with separators.

\paragraph{Substituting}
The popularity of Perl was built on its powerful pattern-matching and substitution features.
The difficulty with substitution is how the API conveys whether one, many, or all occurrences are replaced, and whether a sequence of elements or a single element is replaced.

\paragraph{Testing}
Strings provide many predicates, most importantly determining emptiness, or inclusion of a particular substring, prefix or suffix.
Other predicates range from representation concerns, like determining if all characters belong to the ASCII subset, or of a more ad-hoc nature, like checking if the string is all uppercase or parses as an identifier.  

\paragraph{Iterating}
Strings are often treated as collections of items.
In Pharo a string is a collection of characters and as such it inherits all the high-level iterators defined in \ct{SequenceableCollection} and subclasses.
Similarly, Haskell's \ct{Data.String} is quite terse (just 4 or so functions), but since strings are \ct{Lists}, the whole panoply of higher-level functions on lists are available: \ct{foldr}, \ct{map}, etc.

\paragraph{Endogenous conversion}
Strings can be transformed into other strings according to domain-specific rules:
this covers encoding and escaping, case transpositions, pretty-printing, natural language inflexion, etc.

\paragraph{Exogenous conversion}
Since strings serve as a human-readable representation or serialization format, they can be parsed back into non-string types such as numbers, URLs, or file paths.

\paragraph{Mutating vs copying}
Strings may be considered as collections and provide methods to modify their contents in-place, as opposed to returning a new string with different contents from the original.
Note that this point is orthogonal to the other ones, but influences the design of the whole library.

Mutating strings is dangerous, because strings are often used as value objects, and it is not clear at first sight if a method has side-effects or not.
For example, in \ct{translateToUppercase}, the imperative form hints that it is an in-place modification, but not in \ct{trim}.
Also, safe transformations often rely on their side-effect counterpart: for instance, the safe \ct{asUppercase} sends \ct{translateToUppercase} to a copy of its receiver.

In the case of strings, we believe methods with side effects should be clearly labeled as low-level or private, and their use discouraged; moreover, a clear and systematic naming convention indicating the mutable behavior of a method would be a real plus.
Finally, future developments of the Pharo VM include the Spur object format, which supports immutable instances; this is an opportunity to make literal strings safe\footnote{While clever uses for mutable literals have been demonstrated in the past, we think it is a surprising feature and should not be enabled by default.}, and to reduce copying by sharing character data between strings.

\section{Strings in Other Languages}
\label{sec:others}

To support the analysis and redesign of the current string libraries in Pharo,
we analysed the situation in several other languages.
We took two criteria into account to select the languages below: mainstream
object-oriented languages but also new languages showing alternative designs.
Indeed, our study is about the design of the API at the level of features, how
they compose together and in relation with other types, and how they are
organized in terms of individual methods or functions.
In that light, we believe that the underlying programming paradigm is just one
of many factors that influence the API design.
For instance, it would be possible and probably desirable to have fewer
side-effects and more declarative method names in Pharo's string API, resulting
in a style much closer to functional programming than the current string
implementation; Haskell, with its own limits, provides a worthwile reference
point in that direction.

We will present the key characteristics of the design of strings in Haskell,
Java, Python, Ruby, and Rust.
Then we will discuss some of the used design.

\subsection{Haskell}

In Haskell, the default string implementation \ct{Data.String} is actually a linked list of characters.
This was a design choice to reuse the existing pattern matching and list manipulation functions with virtually no string-specific code; but it is also known to have a huge space overhead and bad performance characteristics for usual string use.
However, if we look further than the core libraries that come with GHC, the Haskell Platform distribution also provides \ct{Data.Text}, an implementation of strings based on a packed array of UTF-16 codepoints.
The same package also includes a lazy variant of that data structure.

In terms of interfaces,
\ct{Data.List}\footnote{\url{https://hackage.haskell.org/package/base-4.9.0.0/docs/Data-List.html}}
and
\ct{Data.Text}\footnote{\url{https://hackage.haskell.org/package/text-1.2.2.1/docs/Data-Text.html}}
are of similar sizes (respectively 116 and 94~functions), but share 60~functions
in common, including \ct{Data.Text.append} and \ct{Data.Text.index} which are
defined as the \ct{(++)} and \ct{(!!)})
operators in \ct{Data.List} (see \tabref{haskellfns}).
This is because many list functions do not apply to lists of characters:
\ct{lookup} expects an association list, \ct{and} \& \ct{or} expect lists of
booleans, \ct{sum} expects a list of numbers, etc.
Conversely, \ct{Data.Text} defines additional functions that are related to
formatting (\ct{center}, \ct{justifyLeft}, \ct{toLower}, \ct{toTitle}), cleaning
up (\ct{dropAround}, \ct{strip}), or parsing text (\ct{breakOn}, \ct{split}), or
parsing text (\ct{breakOn}, \ct{split}).

\begin{table}[tb]\small
  \begin{tabular}{@{}l@{}}
    \toprule
    Haskell--- 60 functions common to both \ct{Data.List} and \ct{Data.Text}:
    \\\footnotesize\sffamily
    \begin{tabularx}{\linewidth}{@{}XXXXX@{}}
      (!!) index   & findIndex & intercalate & minimum     & tail \\
      (++) append  & foldl     & intersperse & null        & tails \\
      all          & foldl'    & isInfixOf   & partition   & take \\
      any          & foldl1    & isPrefixOf  & replicate   & takeWhile \\
      break        & foldl1'   & isSuffixOf  & reverse     & transpose \\
      concat       & foldr     & last        & scanl       & uncons \\
      concatMap    & foldr1    & length      & scanl1      & unfoldr \\
      drop         & group     & lines       & scanr       & unlines \\
      dropWhile    & groupBy   & map         & scanr1      & unwords \\
      dropWhileEnd & head      & mapAccumL   & span        & words \\
      filter       & init      & mapAccumR   & splitAt     & zip \\
      find         & inits     & maximum     & stripPrefix & zipWith \\
    \end{tabularx}
    \\\midrule
    56 functions specific to \ct{Data.List}:
    \\\footnotesize\sffamily
    \begin{tabularx}{\linewidth}{@{}XXXX@{}}
      (\textbackslash\textbackslash)
                       & genericSplitAt  & or           & unzip4 \\
      and              & genericTake     & permutations & unzip5 \\
      cycle            & insert          & product      & unzip6 \\
      delete           & insertBy        & repeat       & unzip7 \\
      deleteBy         & intersect       & scanl'       & zip3 \\
      deleteFirstsBy   & intersectBy     & sort         & zip4 \\
      elem             & isSubsequenceOf & sortBy       & zip5 \\
      elemIndex        & iterate         & sortOn       & zip6 \\
      elemIndices      & lookup          & subsequences & zip7 \\
      findIndices      & maximumBy       & sum          & zipWith3 \\
      genericDrop      & minimumBy       & union        & zipWith4 \\
      genericIndex     & notElem         & unionBy      & zipWith5 \\
      genericLength    & nub             & unzip        & zipWith6 \\
      genericReplicate & nubBy           & unzip3       & zipWith7 \\
    \end{tabularx}
    \\\midrule
    34 functions specific to \ct{Data.Text}:
    \\\footnotesize\sffamily
    \begin{tabularx}{\linewidth}{@{}XXXX@{}}
      breakOn        & count        & snoc        & toCaseFold \\
      breakOnAll     & dropAround   & split       & toLower \\
      breakOnEnd     & dropEnd      & splitOn     & toTitle \\
      center         & empty        & strip       & toUpper \\
      chunksOf       & justifyLeft  & stripEnd    & unfoldrN \\
      commonPrefixes & justifyRight & stripStart  & unpack \\
      compareLength  & pack         & stripSuffix & unpackCString\# \\
      cons           & replace      & takeEnd \\
      copy           & singleton    & takeWhileEnd \\
    \end{tabularx}
    \\\bottomrule
  \end{tabular}
  \caption{Functions defined by Haskell modules \ct{Data.List} and
    \ct{Data.Text}}
  \label{tab:haskellfns}
\end{table}

\subsection{Java}

In Java, instances of the \ct{String} class are immutable (See \tabref{javafns}).
This means that strings can be shared, but also that concatenating them
allocates and copies memory.
To build complex strings while limiting memory churn, the standard library
provides \ct{StringBuilder} and \ct{StringBuffer}; both have the exact same
interface, except the latter is thread-safe.
Finally, \ct{CharSequence} is an interface which groups a few methods for simple
read-only access to string-like objects; it seems like it has a similar purpose
as Rust's slices, but Java strings do not appear to share their underlying
character data: \ct{subSequence()} is the same as \ct{substring()}, which copies
the required range of characters.

Third-party libraries such as Apache
Commons\footnote{\url{https://commons.apache.org}} provide additional
string-related methods in utility classes such as \ct{StringUtils}.
However, since those classes only define static methods, they do not lend
themselves to late binding and polymorphism.

\begin{table}[tb]\small
  \begin{tabular}{@{}l@{}}
    \toprule
    Java--- 35 methods in \ct{String}:
    \\\footnotesize\sffamily
    \begin{tabularx}{\linewidth}{@{}XXXX@{}}
      charAt              & endsWith         & lastIndexOf        & startsWith \\
      codePointAt         & equals           & length             & subSequence \\
      codePointBefore     & equalsIgnoreCase & matches            & substring \\
      codePointCount      & getBytes         & offsetByCodePoints & toCharArray \\
      compareTo           & getChars         & regionMatches      & toLowerCase \\
      compareToIgnoreCase &                  & replace            & toString \\
      concat              & indexOf          & replaceAll         & toUpperCase \\
      contains            & intern           & replaceFirst       & trim \\
      contentEquals       & isEmpty          & split              & hashCode \\
    \end{tabularx}
    \\\midrule
    24 methods in \ct{StringBuffer}/\ct{StringBuilder}:
    \\\footnotesize\sffamily
    \begin{tabularx}{\linewidth}{@{}XXXX@{}}
      append          & codePointCount & insert             & setCharAt \\
      appendCodePoint & delete         & lastIndexOf        & setLength \\
      capacity        & deleteCharAt   & length             & subSequence \\
      charAt          & ensureCapacity & offsetByCodePoints & substring \\
      codePointAt     & getChars       & replace            & toString \\
      codePointBefore & indexOf        & reverse            & trimToSize \\
    \end{tabularx}
    \\\bottomrule
  \end{tabular}
  \caption{Methods defined in Java on string-like classes}
  \label{tab:javafns}
\end{table}

\subsection{Python}

Python's string type is
\ct{str}\footnote{\url{https://docs.python.org/3.6/library/stdtypes.html\#textseq}},
an immutable sequence of Unicode codepoints, whose methods are listed in
\tabref{pythonfns}.
Besides those methods, it also inherits special methods that implement the
behavior for the sequence-related expressions (index-based access, count and
presence of elements).
A few additional functions are defined in module
\ct{string}\footnote{\url{https://docs.python.org/3.6/library/string.html}},
most notably \ct{printf}-style formatting, and Python also provides
\ct{io.StringIO}, a stream-like object to compose large strings efficiently, but
this provides a limited API similar to a file stream, unlike Java's
\ct{StringBuilder} which supports insertion and replace operations.

The general impression is that the API is pretty terse, especially since there
are some symmetric sets of methods, \ie \ct{strip}, \ct{lstrip}, \ct{rstrip}.
Some methods seem too specialized to be present in such a small API (\eg
\ct{swapcase}, \ct{title}, \ct{istitle}).

Finally, since Python, like Ruby, does not have an individual character type,
some character-specific behavior is reported on strings: out of 11~predicates,
only two really apply specifically to strings (\ct{isidentifier} and
\ct{istitle}), the other~9 being universally quantified character predicates.
Encoding and decoding between bytes and Unicode strings is done via the
\ct{str.encode()} and \ct{bytes.decode()} methods, which rely on another
package: \ct{codecs}; here again, character-specific or encoding-specific
behavior does not seem to exist as first-class objects, as codecs are specified
by name (strings).

\begin{table}[tb]\small
  \begin{tabular}{@{}l@{}}
    \toprule
    Python--- 42 methods in \ct{str}:
    \\\footnotesize\sffamily
    \begin{tabularx}{\linewidth}{@{}XXXXXX@{}}
      capitalize & find        & isdigit      & isupper   & rfind      & startswith \\
      casefold   & format      & isidentifier & join      & rindex     & strip \\
      center     & format\_map & islower      & ljust     & rjust      & swapcase \\
      count      & index       & isnumeric    & lower     & rpartition & title \\
      encode     & isalnum     & isprintable  & lstrip    & rstrip     & translate \\
      endswith   & isalpha     & isspace      & partition & split      & upper \\
      expandtabs & isdecimal   & istitle      & replace   & splitlines & zfill \\
    \end{tabularx}
    \\\bottomrule
  \end{tabular}
  \caption{Methods defined in Python on the \ct{str} text sequence type}
  \label{tab:pythonfns}
\end{table}

\subsection{Ruby}

Ruby's strings are mutable sequence of
bytes\footnote{\url{http://www.rubydoc.info/stdlib/core/String}}; however, each
\ct{String} instance knows its own encoding.
Ruby's message send syntax is quite expressive, and many of its APIs make
extensive use of optional parameters and runtime type cases to provide behavior
variants.

A first example is the convention that iteration methods \ct{each\_byte},
\ct{each\_char}, \ct{each\_codepoint}, and \ct{each\_line} either behave as an
internal iterator (\ie a higher-order function) when passed a block, or return
an enumerator object when the block is omitted (external iteration).

A second example is the \ct{[\,]} method, which implements the square bracket
notation for array access; on strings, this is used for substring extraction,
and accepts a number of parameter patterns:
\begin{itemize}
\item a single index, returning a substring of length one (Ruby does not have an
  individual character type),
\item a start index and an explicit length,
\item a range object, locating the substring by start/end bounds instead of by
  its length,
\item a regular expression, optionally with a capture group specifying which
  part of the matched substring to return,
\item another string, returning it if it occurs in the receiver.
\end{itemize} Note also that indices can be negative, in which case they are
relative to the end of the string.

Another widely adopted naming convention in Ruby is that methods with names
terminated by an exclamation point modify their receiver in-place instead of
returning a modified copy; strings are a nice example of this pattern, as more
than a third of the methods belong to such copy/in-place pairs.

\begin{table}[tb]\small
  \begin{tabular}{@{}l@{}}
    \toprule
    Ruby--- 116 methods in \ct{String}:
    \\\footnotesize\sffamily
    \begin{tabularx}{\linewidth}{@{}XXXXX@{}}
      \%                    & codepoints      & initialize  & size \\
      \textasteriskcentered & concat          & replace     & slice (!) \\
      +                     & count           & insert      & split \\
      –                     & crypt           & inspect     & squeeze (!) \\
      <{}<                  & delete (!)      & intern      & start\_with? \\
      <=>                   & downcase (!)    & length      & strip (!) \\
      ==                    & dump            & lines       & sub (!) \\
      ===                   & each\_byte      & ljust       & succ (!) \\
      =\textasciitilde      & each\_char      & lstrip (!)  & sum \\
      {[\,]}                & each\_codepoint & match       & swapcase (!) \\
      {[\,]=}               & each\_line      & next (!)    & to\_c \\
      ascii\_only?          & empty?          & oct         & to\_f \\
      b                     & encode (!)      & ord         & to\_i \\
      bytes                 & encoding        & partition   & to\_r \\
      bytesize              & end\_with?      & prepend     & to\_s \\
      byteslice             & eql?            & replace     & to\_str \\
      capitalize (!)        & force\_encoding & reverse (!) & to\_sym \\
      casecmp               & freeze          & rindex      & tr (!) \\
      center                & getbyte         & rjust       & tr\_s (!) \\
      chars                 & gsub (!)        & rpartition  & unpack \\
      chomp (!)             & hash            & rstrip (!)  & upcase (!) \\
      chop (!)              & hex             & scan        & upto \\
      chr                   & include?        & scrub (!)   & valid\_encoding? \\
      clear                 & index           & setbyte \\
    \end{tabularx}
    \\\bottomrule
  \end{tabular}
  \caption{Methods defined in Ruby's \ct{String} class.
    Methods marked with (!) have an associated in-place version following the
    Ruby naming convention; e.g. \ct{upcase} returns an uppercased copy while
    \ct{upcase!} modifies the receiver in-place.}
  \label{tab:rubyfns}
\end{table}

\subsection{Rust}

Rust has two main types for character strings: \emph{string slices}, represented
by the pointer type
\ct{\&str}\footnote{\url{https://doc.rust-lang.org/std/primitive.str.html}}, and
the boxed type
\ct{String}\footnote{\url{https://doc.rust-lang.org/std/string/struct.String.html}}
(\tabref{rustfns}).
Both types store their contents as UTF-8 bytes; however, while \ct{String} is an
independent object that owns its data, allocates it on the heap and grows it as
needed, \ct{\&str} is a view over a range of UTF-8 data that it does not own
itself.
Literal strings in Rust code are immutable \ct{\&str} slices over
statically-allocated character data.

Making a \ct{String} from a \ct{\&str} slice thus requires allocating a new
object and copying the character data, while the reverse operation is cheap.
In fact, the compiler will implicitly cast a \ct{String} into a \ct{\&str} as
needed, which means that in practice, all methods of slices are also available
on boxed strings, and \ct{String} only adds methods that are concerned with the
concrete implementation.

An surprising design decision in Rust is that strings do \emph{not} implement the array-like indexing operator.
Instead, to access the contents of a string, the library requires explicit use of iterators.
This motivated by the tension between the need, as a systems programming language, to have precise control of memory operations, and the fact that practical, modern encodings (be it UTF-8 or UTF-16) encode characters into a varying number of bytes.
Variable-length encoding makes indexed access to individual characters via dead-reckoning impossible: since the byte index depends on the space occupied by all preceding characters, one has to iterate from the start of the string.
The implications are two-fold: first, this design upholds the convention that array-like indexing is a constant-time operation returning values of fixed size.
Second, multiple iterators are provided on equal footing (methods \ct{bytes()}, \ct{chars()}, \ct{lines()}, or \ct{split()}), each of them revealing a different abstraction level, with no intrinsic or default meaning for what the \emph{n}-th element of a string is; this also makes the interface more uniform.

\begin{table}[tb]\small
  \begin{tabular}{@{}l@{}}
    \toprule
    Rust--- 43 methods defined on string slices \ct{\&str}:
    \\\footnotesize\sffamily
    \begin{tabularx}{\linewidth}{@{}XXXX@{}}
      as\_bytes       & find               & rfind              & splitn \\
      as\_ptr         & into\_string       & rmatch\_indices    & starts\_with \\
      bytes           & is\_char\_boundary & rmatches           & to\_lowercase \\
      char\_indices   & is\_empty          & rsplit             & to\_uppercase \\
      chars           & len                & rsplit\_terminator & trim \\
      contains        & lines              & rsplitn            & trim\_left \\
      encode\_utf16   & match\_indices     & split              & trim\_left\_matches \\
      ends\_with      & matches            & split\_at          & trim\_matches \\
      escape\_debug   & parse              & split\_at\_mut     & trim\_right \\
      escape\_default & replace            & split\_terminator  & trim\_right\_matches \\
      escape\_unicode & replacen           & split\_whitespace \\
    \end{tabularx}
    \\\midrule
    26 methods defined on the boxed \ct{String} type:
    \\\footnotesize\sffamily
    \begin{tabularx}{\linewidth}{@{}XXXX@{}}
      as\_bytes    & from\_utf16\_lossy & is\_empty & reserve \\
      as\_mut\_str & from\_utf8         & len       & reserve\_exact \\
      as\_str      & from\_utf8\_lossy  & new       & shrink\_to\_fit \\
      capacity     & insert             & pop       & truncate \\
      clear        & insert\_str        & push      & with\_capacity \\
      drain        & into\_boxed\_str   & push\_str \\
      from\_utf16  & into\_bytes        & remove \\
    \end{tabularx}
    \\\bottomrule
  \end{tabular}
  \caption{Methods defined in Rust on strings and string slices}
  \label{tab:rustfns}
\end{table}

\section{Reflection on String APIs}
\label{sec:takeaways}

It is difficult to form an opinion on the design of an API before getting
feedback from at least one implementation attempt.  Still, at this stage, we can
raise some high level points that future implementors may consider.  We start
by discussing some issues raised in the analysis of the previous languages,
then we sketch some proposals for a future implementation.

\subsection{Various APIs in Perspective}

While proper assessment of API designs would be more suited for a publication in
cognitive sciences, putting a few languages in perspective during this cursory
examination of the string API raised a few questions.

\paragraph{First-class characters or codepoints}
In Ruby or Python, characters are strings of length one, which has strange implications on some methods, as shown in table~\ref{tab:chrvsstr}.
Was that choice made because the concept of character or codepoint was deemed
useless?
If so, is it due to lack of need in concrete use-cases, or due to early
technical simplifications, technical debt and lack of incentives to change?
If not, is it undertaken by separate encoding-related code, or by strings, even
though it will be often used on degenerate single-character instances?
There is a consensus nowadays around Unicode, which makes encoding conversions a
less pressing issue; however, Unicode comes with enough complexities of its own
---without even considering typography--- that it seems a dedicated
character/codepoint type would be useful.
For instance, Javascript implements strings as arrays of 16-bit integers to be
interpreted as UTF-16, but without taking surrogate sequences into account,
which means that the \ct{length} method is not guaranteed to always return the
actual number of characters in a string.

\begin{table}[tb]\small
  \begin{tabularx}{\linewidth}{@{}>{\small\sffamily}l@{ }>{\small\sffamily}X@{}}
    \toprule
    \rmfamily Python: \\
    ord('a') & $\Rightarrow$ 97 \\
    ord('abc') & TypeError: ord() expected a character, but string of length 3 found \\
    ord('') & TypeError: ord() expected a character, but string of length 0 found \\
    \midrule
    \rmfamily Ruby: \\
    ?a.class & $\Rightarrow$ String \\
    ?a.ord & $\Rightarrow$ 97 \\
    'a'.ord & $\Rightarrow$ 97 \\
    'abc'.ord & $\Rightarrow$ 97 \\
    ''.ord & ArgumentError: empty string \\
    \bottomrule
  \end{tabularx}
  \caption{Character / string confusion in Python and Ruby.
    Both languages use degenerate strings in place of characters;
    Ruby does have a literal character syntax, but it still represents a one-character string.}
  \label{tab:chrvsstr}
\end{table}

\paragraph{Sharing character data}
Second, there is a compromise between expressivity and control over side
effects, data copying, and memory allocation.
Many applications with heavy reliance on strings (\eg parsers, web servers)
benefit from sharing character data across several string instances, because of
gains both in memory space and in throughput; however, this requires that the
shared data does not change.
In this regard, Rust's string slices are interesting because they provide
substrings of constant size and creation time without adding complexity to the
API.
Conversely, Haskell's lazy string compositions, or data structures like ropes,
provide the equivalent for concatenation, without a distinct interface like
Java's \ct{StringBuilder}.

\paragraph{Matching and regular patterns}
Regular patterns, in languages where they are readily available, are highly
effective at analyzing strings.
We did not discuss them here, because while they are a sister feature of
strings, they really are a domain-specific language for working on strings that
can be modularized independently, much like full-fledged parsers.
A way to do that is to make regular expressions polymorphic with other
string-accessing types such as indices, ranges, or strings as patterns); Ruby
does this by accepting various types as argument of its indexing/substring
methods, and Rust by defining a proper abstract type \ct{Pattern} that regular
patterns implement.

\subsection{Concerns for a New String Implementation}

For an API to provide rich behavior without incurring too much cognitive load,
it has to be regular and composable.

\paragraph{Strings and characters are different concepts}
The distinction between character and string types distributes functionality in
adequate abstractions. Characters or codepoints can offer behavior related to
their encoding, or even typographic or linguistic information such as which
alphabet they belong to.

Note that the implementation does not have to be naive and use full-fledged
character objets everywhere.  In Pharo, \ct{String} is implemented as a byte or
word array in a low-level encoding, and \ct{Character} instances are only
created on demand. Most importantly, a character is not a mere limited-range
integer. In this regard, the design of Rust validates that design choice.

\paragraph{Strings are sequences, but not collections}
Strings differ from usual lists or arrays in that containing a specific element
does not really matter \emph{per se}; instead, their contents have to be
interpreted or parsed.
We think this is why their iteration interface is both rich and ad~hoc, and
follows many arbitrary contextual conventions like character classes or
capitalization.
From this perspective, we should probably reconsider the historical design
choice to have \ct{String} be a \ct{Collection} subclass.

\paragraph{Iterations}
Strings represent complex data which can be queried, navigated, iterated in
multiple ways (bytes, characters, words, lines, regular expression matches…).

Iteration based on higher-order functions is an obvious step in this direction;
Smalltalk dialects use internal iterators as the iconic style to express and
compose iterations, but this seems to have discouraged the appearance of an
expressive set of streaming or lazy abstractions like Ruby's enumerators or
Rust's iterators.
Therefore, external iterators should be investigated, under the assumption
that extracting the control flow may lead to better composeability.  Of course,
co-design between strings and collection/stream libraries would be beneficial.

\paragraph{Encodings}
It is misguided to assume that characters always directly map to bytes,
or that any sequence of bytes can be viewed as characters.
To bridge bytes and characters, encodings are required; the API should take them
into account explicitly, including provisions for impossible conversions and
probably for iteration of string contents simultaneously as characters and as
encoded data.

\paragraph{String buffers and value strings}
Pharo strings currently have a single mutable implementation which is used in
two distinct roles: as a value for querying and composing, and as a buffer for
in-place operations.
Streams can assemble large strings efficiently, but more complex editing
operations rely on fast data copying because of the underlying array
representation.

Distinguishing these two roles would allow for internal representations more
suited to each job and for a more focused API.
In particular, the guarantees offered by immutable strings and views like Rust's
slices open many possibilities for reducing data copies and temporary object
allocations.

\paragraph{Consistency and cleanups}

Finally, we would like to consolidate close methods into consistently named
groups or even chains of methods whenever possible.
Immutable strings would favor a declarative naming style.

The current implementation suffers from the presence of many ad-hoc convenience
methods, many of which do not belong in the core API of strings and should be
extracted or removed.

Several methods are related to converting between strings and other kinds of
objects or values.
These conversion methods come in a limited set that is neither generic nor
complete; instead we would prefer a clear, generic, but moldable API for parsing
instances of arbitrary classes out of their string representations.

\section{Discussion and Perspectives}
\label{sec:conclusion}

In this paper, we assess the design of character strings in Pharo.
While strings are simple data structures, their interface is surprisingly large.
Indeed, strings are not simple collections of elements; they can be seen both as explicit sequences of characters, and as simple but very expressive values from the domain of a language or syntax.
In both cases, strings have to provide a spectrum of operations with many intertwined characteristics: abstraction or specialization, flexibility or convenience.
We analyze the domain and the current implementation to identify recurring idioms and smells.

The idioms and smells we list here deal with code readability and reuseability at the level of messages and methods; they fall in the same scope as Kent Beck's list \cite{Beck97a}.
While the paper focuses on strings, the idioms we identify are not specific to strings, but to collections, iteration, or parameter passing; modulo differences in syntax and style usages, they apply to other libraries or object-oriented programming languages.
To identify the idioms and smells, we rely mostly on code reading and the usual tools provided by the Smalltalk environment.
This is necessary in the discovery stage, but it raises several questions:
\begin{itemize}
\item How to document groups of methods that participate in a given idiom?
  As we say in Section~\ref{sec:problem}, method protocols are not suitable: they partition methods by feature or theme, but idioms are overlapping patterns of code factorization and object interaction.
\item How to specify, detect, check, and enforce idioms in the code?
  This is related to architecture conformance techniques \cite{Duca09c}.
\end{itemize}

\bibliographystyle{elsarticle-num}
\bibliography{001_others,002_rmod,003_links}

\begin{thebibliography}{10}
\expandafter\ifx\csname url\endcsname\relax
  \def\url#1{\texttt{#1}}\fi
\expandafter\ifx\csname urlprefix\endcsname\relax\def\urlprefix{URL }\fi
\expandafter\ifx\csname href\endcsname\relax
  \def\href#1#2{#2} \def\path#1{#1}\fi

\bibitem{Blan08a}
J.~Blanchette, The little manual of {API} design,
  \url{http://www4.in.tum.de/~blanchet/api-design.pdf} (Jun. 2008).

\bibitem{Styl06a}
J.~Stylos, S.~Clarke, B.~Myers, Comparing {API} design choices with usability
  studies: A case study and future directions, in: P.~Romero, J.~Good, E.~A.
  Chaparro, S.~Bryant (Eds.), 18th Workshop of the Psychology of Programming
  Interest Group, University of Sussex, 2006, pp. 131--139.
\newblock \href {http://dx.doi.org/10.1.1.102.8525}
  {\path{doi:10.1.1.102.8525}}.

\bibitem{Styl07a}
J.~Stylos, B.~Myers, Mapping the space of {API} design decisions, in: {IEEE}
  Symposium on Visual Languages and Human-Centric Computing, 2007, pp. 50--57.
\newblock \href {http://dx.doi.org/10.1109/VLHCC.2007.44}
  {\path{doi:10.1109/VLHCC.2007.44}}.

\bibitem{Picc13a}
M.~Piccioni, C.~A. Furia, B.~Meyer, An empirical study of {API} usability, in:
  {IEEE/ACM} Symposium on Empirical Software Engineering and Measurement, 2013.
\newblock \href {http://dx.doi.org/10.1109/ESEM.2013.14}
  {\path{doi:10.1109/ESEM.2013.14}}.

\bibitem{Beck97a}
K.~Beck,
  \href{http://stephane.ducasse.free.fr/FreeBooks/BestSmalltalkPractices/Draft-Smalltalk%20Best%20Practice%20Patterns%20Kent%20Beck.pdf}{{Smalltalk}
  Best Practice Patterns}, Prentice-Hall, 1997.
\newline\urlprefix\url{http://stephane.ducasse.free.fr/FreeBooks/BestSmalltalkPractices/Draft-Smalltalk%20Best%20Practice%20Patterns%20Kent%20Beck.pdf}

\bibitem{Cwal05a}
K.~Cwalina, B.~Abrams, Framework Design Guidelines: Conventions, Idioms, and
  Patterns for Reusable {.Net} Libraries, 1st Edition, Addison-Wesley
  Professional, 2005.

\bibitem{Bloch06a}
J.~Bloch, \href{http://doi.acm.org/10.1145/1176617.1176622}{How to design a
  good api and why it matters}, in: Companion to the 21st ACM SIGPLAN Symposium
  on Object-oriented Programming Systems, Languages, and Applications, OOPSLA
  '06, ACM, 2006, pp. 506--507.
\newblock \href {http://dx.doi.org/10.1145/1176617.1176622}
  {\path{doi:10.1145/1176617.1176622}}.
\newline\urlprefix\url{http://doi.acm.org/10.1145/1176617.1176622}

\bibitem{Gris96a}
R.~E. Griswold, M.~T. Griswold,
  \href{http://www.peer-to-peer.com/catalog/language/icon.html}{The Icon
  Programming Language}, Peer-to-Peer Communications, 1996.
\newline\urlprefix\url{http://www.peer-to-peer.com/catalog/language/icon.html}

\bibitem{Berg05a}
A.~Bergel, S.~Ducasse, O.~Nierstrasz, R.~Wuyts,
  \href{http://rmod.inria.fr/archives/papers/Berg05a-CompLangESUG04-classboxesJournal.pdf}{Classboxes:
  Controlling visibility of class extensions}, Journal of Computer Languages,
  Systems and Structures 31~(3-4) (2005) 107--126.
\newblock \href {http://dx.doi.org/10.1016/j.cl.2004.11.002}
  {\path{doi:10.1016/j.cl.2004.11.002}}.
\newline\urlprefix\url{http://rmod.inria.fr/archives/papers/Berg05a-CompLangESUG04-classboxesJournal.pdf}

\bibitem{Clif00a}
C.~Clifton, G.~T. Leavens, C.~Chambers, T.~Millstein, {MultiJava}: Modular open
  classes and symmetric multiple dispatch for {Java}, in: {OOPSLA} 2000
  Conference on Object-Oriented Programming, Systems, Languages, and
  Applications, 2000, pp. 130--145.

\bibitem{Wool98a}
B.~Woolf, Null object, in: R.~Martin, D.~Riehle, F.~Buschmann (Eds.), Pattern
  Languages of Program Design 3, Addison Wesley, 1998, pp. 5--18.

\bibitem{Mure96a}
S.~Murer, S.~Omohundro, D.~Stoutamire, C.~Szyperski, Iteration abstraction in
  sather, {ACM} Transactions on Programming Languages and Systems 18~(1) (1996)
  1--15.
\newblock \href {http://dx.doi.org/10.1145/225540.225541}
  {\path{doi:10.1145/225540.225541}}.

\bibitem{transducers}
R.~Hickey, Clojure transducers, \url{http://clojure.org/transducers}.

\bibitem{ANSI98a}
ANSI, New York, {A}merican {N}ational {S}tandard for {I}nformation {S}ystems --
  {P}rogramming {L}anguages -- {S}malltalk, ANSI/INCITS 319-1998,
  \url{http://wiki.squeak.org/squeak/uploads/172/standard\_v1\_9-indexed.pdf}
  (1998).

\bibitem{Duca09c}
S.~Ducasse, D.~Pollet,
  \href{http://rmod.inria.fr/archives/papers/Duca09c-TSE-SOAArchitectureExtraction.pdf}{Software
  architecture reconstruction: A process-oriented taxonomy}, IEEE Transactions
  on Software Engineering 35~(4) (2009) 573--591.
\newblock \href {http://dx.doi.org/10.1109/TSE.2009.19}
  {\path{doi:10.1109/TSE.2009.19}}.
\newline\urlprefix\url{http://rmod.inria.fr/archives/papers/Duca09c-TSE-SOAArchitectureExtraction.pdf}

\end{thebibliography}

\appendix
\section*{Appendix --- Classifying the Pharo String API}
\label{sec:classification}

\subsubsection*{Finding}
Methods returning places in the string (indices, ranges).

\begin{tabbing}
  \hspace{.5\linewidth}\= \kill
  \ct{findString:}
  \> \ct{findString:startingAt:} \\
  \ct{findString:startingAt:caseSensitive:} \\
  \ct{findLastOccurrenceOfString:startingAt:} \\
  \ct{allRangesOfSubString:}
  \> \ct{findAnySubStr:startingAt:} \\
  \ct{findCloseParenthesisFor:}
  \> \ct{findDelimiters:startingAt:} \\
  \ct{findWordStart:startingAt:} \`{} no senders \\
  \ct{findIn:startingAt:matchTable:} \`{} auxiliary method \\
  \ct{findSubstring:in:startingAt:matchTable:} \`{} auxiliary method \\
  \ct{findSubstringViaPrimitive:in:startingAt:matchTable:} \`{} one sender \\[\smallskipamount]

  \ct{indexOf:}
  \qquad \ct{indexOf:startingAt:}
  \qquad \ct{indexOf:startingAt:ifAbsent:} \\
  \ct{indexOfSubCollection:} \`{} mispackaged \\
  \ct{indexOfSubCollection:startingAt:ifAbsent:} \\
  \ct{indexOfFirstUppercaseCharacter} \`{} redundant, one sender \\
  \ct{indexOfWideCharacterFrom:to:} \\
  \ct{lastSpacePosition} \\[\smallskipamount]

  \ct{lastIndexOfPKSignature:} \`{} adhoc or mispackaged \\[\smallskipamount]

  \ct{skipAnySubStr:startingAt:}
  \> \ct{skipDelimiters:startingAt:}
\end{tabbing}

\subsubsection*{Extracting}
Methods returning particular substrings.

\begin{tabbing}
  \ct{wordBefore:} \\
  \ct{findSelector} \`{} mispackaged, specific to code browser \\
  \ct{findTokens:} \\
  \ct{findTokens:escapedBy:} \`{} no senders (besides tests)\\
  \ct{findTokens:includes:} \`{} one sender \\
  \ct{findTokens:keep:} \\

  \ct{lineCorrespondingToIndex:} \\
  \ct{squeezeOutNumber} \`{} ugly parser, one sender \\

  \ct{splitInteger} \`{} what is the use-case? \\
  \ct{stemAndNumericSuffix} \`{} duplicates previous method
\end{tabbing}

\subsubsection*{Splitting}
Methods returning a collection of substrings.

\begin{tabbing}
  \ct{lines} \\

  \ct{subStrings:} \\
  \ct{substrings} \`{} not a call to previous one, why? \\
  \ct{findBetweenSubStrs:} \\
  \ct{keywords} \`{} adhoc, assumes receiver is a selector
\end{tabbing}

\subsubsection*{Enumerating}

\begin{tabbing}
  \ct{linesDo:}
  \qquad \ct{lineIndicesDo:}
  \qquad \ct{tabDelimitedFieldsDo:}
\end{tabbing}

\subsubsection*{Conversion to other objects}

Many core classes such as time, date and duration that have a compact and meaningful textual description extend the class String to offer conversion from a string to their objects.
Most of them could be packaged with the classes they refer to, but splitting a tiny core into even smaller pieces does not make a lot of sense, and there are legitimate circular dependencies in the core: a string implementation cannot work without integers, for example.
Therefore, most of these methods are part of the string API from the core language point of view:

\begin{tabbing}
  \hspace{.3\linewidth}\= \hspace{.3\linewidth}\= \hspace{.2\linewidth}\=\kill
  \ct{asDate}
  \> \ct{asNumber}
  \> \ct{asString}
  \> \ct{asSymbol} \\

  \ct{asTime}
  \> \ct{asInteger}
  \> \ct{asStringOrText} \\

  \ct{asDuration}
  \> \ct{asSignedInteger}
  \> \ct{asByteArray} \\

  \ct{asDateAndTime}
  \> \ct{asTimeStamp}
\end{tabbing}

Some other methods are not as essential:
\begin{tabbing}
  \ct{asFourCode}\qquad
  \ct{romanNumber}\qquad
  \ct{string}\qquad
  \ct{stringhash}
\end{tabbing}

\subsubsection*{Conversion between strings}

A different set of conversion operations occurs between strings themselves.
\begin{itemize}
\item typography and natural language:
  \ct{asLowercase}, \ct{asUppercase}, \ct{capitalized}, \ct{asCamelCase}, \ct{withFirstCharacterDownshifted},
  \ct{asPluralBasedOn:},
  \ct{translated}, \ct{translatedIfCorresponds}, \ct{translatedTo:}
\item content formatting:
  \ct{asHTMLString}, \ct{asHex}, \ct{asSmalltalkComment}, \ct{asUncommentedSmalltalkCode},
\item internal representation:
  \ct{asByteString}, \ct{asWideString}, \ct{asOctetString}
\end{itemize}

\subsubsection*{Streaming}

\begin{tabbing}
  \ct{printOn:}\qquad
  \ct{putOn:}\qquad
  \ct{storeOn:}
\end{tabbing}

\subsubsection*{Comparing}

\begin{tabbing}
  \ct{caseInsensitiveLessOrEqual:}\quad
  \= \ct{caseSensitiveLessOrEqual:} \\
  \ct{compare:with:collated:}
  \> \ct{compare:caseSensitive:} \\
  \ct{compare:}
  \> \ct{sameAs:}
\end{tabbing}

\subsubsection*{Testing}

\begin{tabbing}
  \hspace{.3\linewidth}\= \hspace{.3\linewidth}\= \kill
  \ct{endsWith:}
  \> \ct{endsWithAnyOf:} \\
  \ct{startsWithDigit}
  \> \ct{endsWithDigit}
  \> \ct{endsWithAColon} \`{} ad-hoc \\
  \ct{hasContentsInExplorer} \`{} should be an extension \\
  \ct{includesSubstring:caseSensitive:}
  \>\> \ct{includesSubstring:} \\
  \ct{includesUnifiedCharacter}
  \>\> \ct{hasWideCharacterFrom:to:} \\
  \ct{isAllDigits}
  \> \ct{isAllSeparators}
  \> \ct{isAllAlphaNumerics} \\
  \ct{onlyLetters} \`{} inconsistent name \\
  \ct{isString}
  \> \ct{isAsciiString}
  \> \ct{isLiteral} \\
  \ct{isByteString}
  \> \ct{isOctetString}
  \> \ct{isLiteralSymbol}\\
  \ct{isWideString} \\
  \ct{beginsWithEmpty:caseSensitive:} \`{} bad name, duplicate \\
  \ct{occursInWithEmpty:caseSensitive:} \`{} bad name, mispackaged
\end{tabbing}

\subsubsection*{Querying}

\begin{tabbing}
  \hspace{.3\linewidth}\= \hspace{.3\linewidth}\= \kill
  \ct{lineCount}
  \> \ct{lineNumber:} \\
  \ct{lineNumberCorrespondingToIndex:}
  \>\> \ct{leadingCharRunLengthAt:} \\
  \ct{initialIntegerOrNil}
  \> \ct{numericSuffix}
  \> \ct{indentationIfBlank:} \\
  \ct{numArgs} \`{} selector-related \\
  \ct{parseLiterals} \`{} contents of a literal array syntax
\end{tabbing}

\subsubsection*{Substituting}

\begin{tabbing}
  \hspace{.4\linewidth}\= \hspace{.2\linewidth}\= \kill
  \ct{copyReplaceAll:with:asTokens:}
  \>\> \ct{copyReplaceTokens:with:} \\

  \ct{expandMacros}
  \> \ct{expandMacrosWithArguments:} \\
  \ct{expandMacrosWith:}
  \> \ct{expandMacrosWith:with:} \\
  \ct{expandMacrosWith:with:with:} \\
  \ct{expandMacrosWith:with:with:with:} \\

  \ct{format:} \\

  \ct{replaceFrom:to:with:startingAt:} \`{} primitive \\

  \ct{translateWith:}
  \> \ct{translateFrom:to:table:} \\
  \ct{translateToLowercase}
  \> \ct{translateToUppercase}
\end{tabbing}

\subsubsection*{Correcting}

\begin{tabbing}
  \ct{correctAgainst:}\qquad
  \ct{correctAgainst:continuedFrom:} \\
  \ct{correctAgainstDictionary:continuedFrom:} \\
  \ct{correctAgainstEnumerator:continuedFrom:}
\end{tabbing}

\subsubsection*{Operations}

\begin{tabbing}
  \hspace{.5\linewidth}\= \kill
  \ct{contractTo:}
  \> \ct{truncateTo:} \\
  \ct{truncateWithElipsisTo:} \\

  \ct{encompassLine:}
  \> \ct{encompassParagraph:} \\

  \ct{withNoLineLongerThan:} \\
  \ct{withSeparatorsCompacted}
  \> \ct{withBlanksCondensed} \\

  \ct{withoutQuoting} \\
  \ct{withoutLeadingDigits}
  \> \ct{withoutTrailingDigits} \\
  \ct{withoutPeriodSuffix}
  \> \ct{withoutTrailingNewlines} \\

  \ct{padLeftTo:}
  \> \ct{padLeftTo:with:} \\
  \ct{padRightTo:}
  \> \ct{padRightTo:with:} \\
  \ct{padded:to:with:} \`{} duplicates the two previous \\

  \ct{surroundedBy:}
  \> \ct{surroundedBySingleQuotes} \\

  \hspace{.3\linewidth}\= \hspace{.3\linewidth}\= \hspace{.3\linewidth}\= \kill
  \ct{trimLeft:right:}
  \> \ct{trim}
  \> \ct{trimmed} \\
  \ct{trimLeft}
  \> \ct{trimBoth}
  \> \ct{trimRight} \\
  \ct{trimLeft:}
  \> \ct{trimBoth:}
  \> \ct{trimRight:}
\end{tabbing}

\subsubsection*{Encoding}

\begin{tabbing}
  \hspace{.5\linewidth}\= \kill
  \ct{convertFromEncoding:}
  \> \ct{convertFromWithConverter:} \\
  \ct{convertToEncoding:}
  \> \ct{convertToWithConverter:} \\

  \ct{convertToSystemString}
  \> \ct{encodeDoublingQuoteOn:} \\

  \ct{withLineEndings:}
  \> \ct{withSqueakLineEndings} \\
  \ct{withUnixLineEndings}
  \> \ct{withInternetLineEndings} \\
  \ct{withCRs} \`{} convenience, used a lot
\end{tabbing}

\subsubsection*{Matching}

\begin{tabbing}
  \ct{alike:}
  \qquad\= \ct{howManyMatch:} \`{} similarity metrics \\
  \ct{charactersExactlyMatching:} \`{} bad name: common prefix length \\

  \ct{match:}
  \> \ct{startingAt:match:startingAt:}
\end{tabbing}

\subsubsection*{Low-Level Internals}

\begin{tabbing}
  \hspace{.3\linewidth}\= \hspace{.3\linewidth}\= \hspace{.3\linewidth}\= \kill
  \ct{hash}
  \> \ct{typeTable} \\

  \ct{byteSize}
  \> \ct{byteAt:}
  \> \ct{byteAt:put:} \\
  \ct{writeLeadingCharRunsOn:}
\end{tabbing}

\subsubsection*{Candidates for removal}

While performing this analysis we identified some possibly obsolete methods.

\begin{tabbing}
  \hspace{.3\linewidth}\= \hspace{.3\linewidth}\= \hspace{.3\linewidth}\= \kill
  \ct{asPathName}
  \> \ct{asIdentifier:}
  \> \ct{asLegalSelector} \\
  \ct{do:toFieldNumber:} \\
  \ct{indexOfFirstUppercaseCharacter}
\end{tabbing}

\end{document}